%Paper: hep-ph/9312254  (Phys. Rev. D 50, 2806 (1994))
%From: carroll@marie.mit.edu (Sean Carroll)
%Date: Thu, 9 Dec 1993 16:49:14 -0500
%Date (revised): Thu, 16 Jun 1994 14:34:11 -0400
%Date (revised): Tue, 25 Jul 1995 16:24:41 -0400

\magnification=1200
\vbadness=10000

\abovedisplayskip=25pt plus 4pt minus 10pt
\abovedisplayshortskip=20pt plus 4pt
\belowdisplayskip=25pt plus 4pt minus 10pt
\belowdisplayshortskip=28pt plus 4pt minus 4pt
\hfuzz=10pt \overfullrule=0pt
\baselineskip=12pt
\parindent 20pt \parskip 6pt
\def\Tr{\mathop{\rm Tr}\nolimits}
\def\det{\mathop{\rm det}\nolimits}
\def\mapright#1{\smash{\mathop{\longrightarrow}\limits^{#1}}}
\def\mapdown#1{\big\downarrow \rlap{$\vcenter
  {\hbox{$\scriptstyle#1$}}$}}
\def\su#1{{\rm SU}(#1)}
\def\so#1{{\rm SO}(#1)}
\def\sp#1{{\rm Sp}(#1)}
\def\u#1{{\rm U}(#1)}
\def\o#1{{\rm O}(#1)}
\def\p#1{{\pi_{#1}}}
\def\Z{{\bf Z}}
\def\M{{\cal M}}

\hfill{~}
\vskip .3in

\centerline{{\bf A TEXTURE BESTIARY}\footnote{$^*$}{This
work was supported in part by NASA under Grants no. NAGW-931 and
NGT-50850, by the National Science Foundation under grants
AST/90-05038 and PHY/92-06867, and by the U.S. Department
of Energy (D.O.E.) under contract no. DE-AC02-76ER03069.}}
\vskip .3in

\centerline {James A. Bryan,$^{(1)}$ Sean M. Carroll,$^{(2)}$ and
Ted Pyne$^{(3)}$}
\vskip .3cm
\centerline{\it $^{(1)}$Department of Mathematics, Harvard University}
\centerline{\it Cambridge, Massachusetts\quad 02138}
\centerline{\it email: gruad@zariski.harvard.edu}
\vskip .3cm
\centerline{\it $^{(2)}$Center for Theoretical Physics, Laboratory for
Nuclear Science}
\centerline{\it and Department of Physics}
\centerline{\it Massachusetts Institute of Technology}
\centerline{\it Cambridge, Massachusetts\quad 02139}
\centerline{\it email: carroll@marie.mit.edu}
\vskip .3cm
\centerline{\it $^{(3)}$Harvard-Smithsonian Center for Astrophysics}
\centerline{\it Cambridge, Massachusetts\quad 02138}
\centerline{\it email: pyne@cfa.harvard.edu}

\vskip .3in

\centerline{\bf Abstract}
\vskip .1in

Textures are topologically nontrivial field configurations which can
exist in a field theory in which a global symmetry group $G$
is broken to a subgroup $H$, if the
third homotopy group $\p3$ of $G/H$ is nontrivial.  We compute this
group for a variety of choices of $G$ and $H$, revealing what
symmetry breaking patterns can lead to texture.  We also comment
on the construction of texture configurations in the different
models.

\vskip .3in

\centerline{Subtitle ``From symmetry-breaking patterns}
\centerline{to topological field configurations'' added in journal.}

\vfill

\centerline{CTP~\# 2263 \hfill December 1993, revised June 1994}
\centerline{hep-ph/9312254\hfill Final Version}
\vskip .1in

\eject

\baselineskip 14pt

\noindent{\bf I. Introduction}

Longstanding interest in topological defects such as vortices
(cosmic strings) and monopoles has recently
been joined by interest in textures, field
configurations which are everywhere at the minimum of the potential
energy of a theory but cannot be smoothly deformed to a constant
configuration without leaving the vacuum manifold.  The cosmological
implications of textures have been extensively studied [1,2,3], as well
as their appearance in ordered media [4].  In field theories with
higher-derivative interactions, texture configurations can lead to
stable solitons such as Skyrmions [5].

Although textures are conceptually distinct from defects, the
computations which reveal their existence are entirely analogous; both
depend on the topology of the vacuum manifold $\M$, the set of
minima of the potential energy in the theory under consideration.
A topological defect will necessarily exist if, on a certain
region of space, the fields take values in $\M$ such that it is
impossible to find a smooth solution over all of space which does not
leave $\M$.  For example, we may be given the value of the field on a
circle $S^1$ in space.  If it is impossible to smoothly deform the
image of this circle in $\M$ to a point, then the field must climb out
of $\M$ somewhere inside the circle, and we know that a vortex must
pass through the circle.  Clearly, such vortices can exist if the
first homotopy group $\p1(\M)$, the set of topologically
inequivalent maps from $S^1\rightarrow\M$, is nontrivial.  Similarly,
monopoles are related to $\p2(\M)$ (topologically inequivalent maps
of two-spheres into $\M$), and domain walls are related to $\p0(\M)$
(topologically inequivalent maps of zero-spheres, or simply the number
of disconnected pieces into which $\M$ falls).

Textures, on the other hand, are field configurations which do remain
in the vacuum manifold everywhere in space, but are nevertheless
topologically distinct from the ``global vacuum'' in which the field
has no gradient energy and lies at the same point of $\M$ everywhere.
(It follows that textures are only important in theories with
spontaneously broken {\it global} symmetries; in a gauge theory, a
``texture'' configuration is merely a gauge transform of a
constant-field configuration.)
A region of space with a spherical boundary may be said to contain a
texture if the field lies in $\M$ at every point,
takes the same value everywhere on the boundary
and is topologically distinct from a constant-field configuration.
Since the boundary maps to a single point in field space, such a
configuration defines a map from a three-sphere into the vacuum
manifold.   Thus, the existence of textures is predicated
on the existence of topologically nontrivial
maps $S^3\rightarrow\M$, which are classified by the third homotopy
group $\p3(\M)$.  For a theory in which a symmetry group $G$ is
spontaneously broken to a subgroup $H$, the vacuum manifold $M$
is isomorphic to $G/H$.  Hence, $\p3(\M)$ can be calculated once
$G$ and $H$ are specified.\footnote{$^1$}{The manifold $G/H$
depends not only on $G$ and $H$, but on the specific embedding of
$H$ in $G$.  In the course of the paper we will specify how
$H$ acts as a subgroup of $G$ for each of the models we
consider.}

It is not difficult to discover symmetry breaking patterns which
lead to texture.  For example, $\so4/\so3 =S^3$,
and it is well known that $\p3(S^3)={\bf Z}$.
However, it would be interesting to know when any specified field
theory predicts texture.  In this paper, we undertake
the task of computing $\p3(G/H)$ for a variety of symmetry-breaking
patterns, thereby constructing a catalogue of when and in what forms
texture can appear.  By the standards of modern algebraic topology
this is not a sophisticated problem; from the point of view of the
practicing cosmologist, however, it is is useful to know both what
the relevant homotopy groups are, and also how field configurations
representing textures may be constructed in a given theory.  We have
therefore endeavored to calculate the topological properties of the
vacuum manifolds in such a way that the discussion would lead
directly to the examination of specific field configurations.

We would be remiss if we failed to mention that the texture scenario
for cosmological structure formation has been dealt
blows on both theoretical and observational grounds.  On the
theoretical front, it has been noticed that symmetry-breaking
operators induced by quantum gravity can drastically alter the
evolution of would-be textures, rendering them cosmologically impotent
[6].  (At the same time, our understanding of Planck-scale physics
is insufficient to make incontrovertible statements about such
effects.)  Observationally, the texture scenario for structure formation,
when normalized to the amplitude of microwave background fluctuations
observed by COBE, predicts galaxy velocity dispersions somewhat
higher than those observed [3].  (Once again, however, our
understanding of galaxy-scale physics is also insufficient to
make incontrovertible statements at this time.)
Finally, it is not clear to what extent the
topologically nontrivial nature of textures is relevant;
investigations have shown that interesting perturbations can result
from scalar field gradients even when $\p3(\M)$=0 [7].

Nevertheless, the resilient nature of cosmological models and the
possible appearance of texture in other contexts suggests to
us that our results are still interesting.  Furthermore, in the
absence of symmetry-breaking operators, a theory which might predict
texture must now be shown to be consistent with the large-scale
structure data, and the
computations done below help to make this possible.

\vskip .5cm
\noindent{\bf II. Basic Techniques}

{\bf A. Set-up}

In this section we describe some general techniques
used to compute $\p3(G/H)$ for various
choices of symmetry breaking patterns.  We review some basic homotopy
lore, including a description of the exact sequence, and
describe some simplifications which occur
when $H$ is abelian.

We first recall the homotopy groups of various spaces,
including the Lie groups in which we are interested.  This information
can be found in reference books, and we have compiled those groups
which are relevant to us in Table~I.  It is also useful to know
that the homotopy of a product of two spaces is simply the sum of
the individual homotopy groups,
  $$\p{q}(X\times Y)=\p{q}(X)\oplus\p{q}(Y).\eqno(2.1)$$
Finally, we shall make use of the fact that $\su{n}/\su{n-1}$
and $\so{n}/\so{n-1}$ are homeomorphic to spheres:
  $$\eqalign{\su{n}/\su{n-1}&\sim S^{2n-1}\cr
  \so{n}/\so{n-1} &\sim S^{n-1}.\cr }\eqno(2.2)$$
Notice that, since $\su1$ and $\so1$ are both the trivial group,
$\su2\sim S^3$ and $\so2\sim S^1$.
The usefulness of these relations stems from the simple nature
of the lower homotopy groups of the spheres:
  $$\p{q}(S^n)=\cases{0 & for $q<n$ ,\cr
  \Z & for $q=n$ .\cr}\eqno(2.3)$$
For $q>n$ there is no easy relation analogous to (2.3); in fact,
the computation of $\p{q}(S^n)$ is an open problem.

{\bf B. Exact Homotopy Sequence}

The exact homotopy sequence
is a sequence of maps between homotopy groups of two
spaces and those of their quotient space.  If $G$ is a Lie group with
subgroup $H$, we have
  $$\ldots\ \mapright{}\ \p{q+1}(G/H)\
  \mapright{\gamma_{q+1}}\ \p{q}(H)\ \mapright{\alpha_q}
  \ \p{q}(G)\ \mapright{\beta_q}\ \p{q}(G/H)
  \ \mapright{\gamma_q}\ \p{q-1}(H)\ \mapright{}\
  \ldots  \eqno(2.4)$$
The maps $\alpha_i$, $\beta_i$ and $\gamma_i$ are specified in terms
of the spaces $G$, $H$ and $G/H$, and they are all group
homomorphisms.  For example, the map $\alpha_q$ takes the image of a
$q$-sphere in $H$ into an image of a $q$-sphere in $G$ using the
inclusion of $H$ as a subgroup in $G$, $i:H\hookrightarrow G$.  For
more details see [8,9].

``Exactness'' means that the image
of each map is precisely equal to the kernel (the set of elements
taken to zero) of the map following it.  An important consequence of
exactness is that if two spaces $A$ and $B$ are sandwiched between the
trivial group, $0\rightarrow A\ \mapright\phi
\ B\rightarrow 0$, then the map $\phi$ must be an isomorphism.
This is easy to see: since the kernel of
$B\rightarrow 0$ is all of $B$, $\phi$ must be onto.  Meanwhile, since
the kernel of $\phi$ is the image of $0\rightarrow A$ (which is just
zero), in order for $\phi$ to be a group homomorphism it must be
one-to-one.  Thus, $\phi$ is an isomorphism.

The exact homotopy sequence can also be used when the space $G$
appearing in (2.4) is not a Lie group, as long as $G$ may be
thought of as a fiber bundle with fiber $H$ and base space $G/H$.
For example, if $H$ is a group which acts freely (only the
identity has fixed points) on a manifold $M$, then $M$ may be
thought of as a fiber bundle with base space $M/H$.  The usefulness
of this fact arises when a global symmetry group $G$ breaks
spontaneously to a subgroup of the form $H_1\times H_2$.  Then we may
think of $G/(H_1\times H_2)$ as $(G/H_1)/H_2$, even if $G/H_1$ is not
a group, since the action of $H_2$ divides $G/H_1$ into
well-defined equivalence classes.  Thus, we can apply (2.4) with
$G/H_1$ playing the role of $G$ and $H_2$ playing the role of $H$.
It is important, however, not to get carried away; if a subgroup $H$
of $G$ can be written in the form $A/B$, it is in no way permissible to
think of $G/(A/B)$ as $(G\times B)/A$.

In computing $\p3(G/H)$, we take advantage of the fact that, for any
Lie group $H$, $\p2(H)=0$.  We therefore consider
  $$\matrix{\p3(H)&\mapright{\alpha}&\p3(G)&\mapright{\beta}&
    \p3(G/H)&\mapright{\gamma}& 0\ .\cr}\eqno(2.5)$$
The map $\gamma$ takes all of $\p3(G/H)$ to zero.  Since the
sequence is exact, the image of $\beta$ is therefore all of
$\p3(G/H)$; {\it i.e.}~$\beta$ is onto.
This implies that $\p3(G/H)$ is isomorphic to the
domain of $\beta$ (which is all of $\p3(G)$) modulo the kernel of
$\beta$ (which we know is ${\rm Im\ }\alpha$).  Thus,
  $$\p3(G/H)\cong \p3(G)/{\rm Im\ }\alpha\ .\eqno(2.6)$$
Therefore, our task is reduced to computing $\p3(G)$ (usually easy to
do) and Im~$\alpha$ (sometimes hard to do).

{\bf C. Abelian Subgroups}

Consider the case of a general global symmetry
group $G$ spontaneously breaking down to an abelian subgroup $A$.
All compact, connected,
abelian Lie groups are of the form $A={\rm U}(1)\times {\rm
U}(1)\times \ldots \times {\rm U}(1)$; disconnected groups will be
of this form times various factors of $\Z$ and $\Z_n$.  Using the
product rule (2.1), along with $\p3({\rm U}(1))=\p3({\Z})=
\p3({\Z_n})=0$, we have $\p3(A)=0$ for any abelian group $A$.
(The compactness condition is actually unnecessary.)  As
Turok [1] has pointed out, the exact sequence
  $$\matrix{\p3(A)&\mapright{}&\p3(G)&\mapright{}&
  \p3(G/A)&\mapright{}&\p2(A)\cr \Vert&&&&&&\Vert\cr
  0&&&&&&0\cr}\eqno(2.7)$$
then implies that
  $$\p3(G/A)=\p3(G),\qquad A{\rm \ abelian}\ .\eqno(2.8)$$
This includes, of course, the case $A=0$.  Further,
when a subgroup $H$ can be decomposed into $H=K\times A$, with $K$
an arbitrary Lie group and $A$ abelian, we can use the reasoning
of Sec.~II.B to show that
  $$\p3(G/(K\times A))=\p3(G/K),\qquad A{\rm \ abelian}\ .\eqno(2.9)$$
The interesting examples in which $H$ is non-abelian must be
handled on a case-by-case basis.

\vskip .5cm
\noindent{\bf III. Simple Groups}

{\bf A. General Procedure}

In this section we present a formula which can be combined with
(2.6) to compute $\p3(G/H)$ when $G$ and $H$ are both
simple.\footnote{$^2$}{We are very grateful to Sidney Coleman
for explaining to us the procedure outlined in this section.}
(Generalization to non-simple groups is straightforward.)
The important property of simple groups is that
$\p3(G)={\bf Z}$ for any simple group $G$;
the element of ${\bf Z}$ to which a map $g:S^3
\rightarrow G$ corresponds is called the winding number of $g$, or
$\eta_G(g)$.  (We denote the range space of $g$ as a subscript for
clarity.)  The map $\alpha:\p3(H)\rightarrow\p3(G)$ whose
image we wish to compute is a homomorphism from
${\bf Z}$ to ${\bf Z}$,
which may be thought of as multiplication by an integer $p$.
In other words, a map $f:S^3\rightarrow H$ with winding number
$\eta_H(f)$ will, when composed with the inclusion map
$i:H\hookrightarrow G$, yield a map with winding number
  $$\eta_G(i\circ f)=p\eta_H(f)\ .\eqno(3.1)$$
Hence, the image of $\alpha$ is every $p$th integer, so (2.6)
and $\p3(G)={\bf Z}$ together imply that
  $$\p3(G/H)={\bf Z}_p\ ,\eqno(3.2)$$
the group of integers modulo addition by $p$.  In the case $p=0$ this
is interpreted as $\Z_0=\Z$, while $\Z_1$ is the trivial
group.  Our aim is therefore to compute $p$.

Clearly, it should be possible to compute $p$ by choosing a generator
$f$ of $\p3(H)$ (a map with unit winding number, $\eta_H(f)=1$);
then (3.1) implies that $p$ is equal to the winding number of
$i\circ f$ in $G$.  Rather than computing this winding number
directly, we shall define a number $\xi_G(\phi)$ which is proportional
to the winding number of a map $\phi$, and then take the ratio of
$\xi_G(i\circ f)$ to the equivalent expression for a map with winding
number one.  In other words, since $p=\eta_G(i\circ f)\propto
\xi_G(i\circ f)$ for a map $f$ which generates
$\p3(H)$, we shall choose a map
$g:S^3\rightarrow G$ which generates $\p3(G)$ (thus,
$\eta_G(g)=1$), and calculate the ratio
  $$p={{\xi_G(i\circ f)}\over{\xi_G(g)}}\ .\eqno(3.3)$$
To do so, we take advantage
of a remarkable fact unique to $\p3$ (as opposed to the other
$\p{q}$).  Given a map $\phi:S^3\rightarrow G$
which takes a point labled
by $x$ to a matrix $\phi(x)$, there is an integral expression
proportional to the winding number:
  $$\eta_G(\phi)\propto \int_{S^3}d^3x\ \epsilon^{ijk}\Tr\left[
  (\partial_i \phi)\phi^{-1}(\partial_j \phi)\phi^{-1}
  (\partial_k \phi)\phi^{-1}\right] \ ,\eqno(3.4)$$
where matrix multiplication is implicit.  A demonstration
of the topological invariance of this quantity can be found in
[10].  Thus, $p$ will follow from Eq.~(3.3) if we can evaluate
(3.4) for the maps $i\circ f$ and $g$.

Fortunately it is not even necessary to evaluate (3.4), as we can
use another remarkable fact unique to $\p3$.  This is a
theorem due to Bott [11], that a map $g$ representing the
generator of $\p3(G)$ may always be taken to be a homomorphism
from $\su2$ to $G$, while maps with winding number $n$ may be
represented by $\phi = g^n$.  The integral
in (3.4) will then be equal to a constant factor
(proportional to the volume of $\su2$, but independent of $\phi$)
times the value of the integrand at any point.  We may take this
point to be the identity element of $\su2$ (with coordinates $x=0$),
which is mapped to the identity element $e$ in $G$.  Thus, (3.4) is
equivalent to
  $$\eta_G(\phi)\propto \epsilon^{ijk}\Tr\left[\partial_i\phi(x)
  \partial_j\phi(x)\partial_k\phi(x)\right]\big\vert_{x=0}\ ,
  \eqno(3.5)$$
where we have used $\phi(0)^{-1}=e^{-1}=e$.  Now, the quantity
$(\partial_i\phi(x))\phi^{-1}(x)$ evaluated at $x=0$ defines an
element of the Lie algebra of $G$. Since the
group map $\phi$ defines an algebra map
  $$\eqalign{\phi_*: {\cal SU}(2)&\rightarrow{\cal G}\cr
  X_i&\mapsto \widetilde X_i\ ,\cr}  \eqno(3.6)$$
(3.5) takes the form
  $$\eta_G(\phi)\propto \epsilon^{ijk} \Tr
  \left(\widetilde X_i\widetilde X_j\widetilde X_k\right)\ .\eqno(3.7)$$
The matrices $\widetilde X_i$ are elements of ${\cal G}$, the Lie
algebra of $G$, but by themselves they for a basis for
${\cal SU}(2)$, the Lie algebra of $\su2$.  Specifically, they satisfy
commutation relations
  $$[\widetilde X_i,\widetilde X_j]=
  n\epsilon^{ijk}\widetilde X_k\ ,\eqno(3.8)$$
from which it follows immediately that
  $$\epsilon^{ijk}\widetilde X_i\widetilde X_j=2n\widetilde X_k\ ,
  \eqno(3.9)$$
where $n$ is the winding number of the map $\phi$.  (If $\phi$
is the generator of $\p3(G)$, we have $n=1$, and $\phi_*$ is a
Lie algebra homomorphism.)
Thus, (3.7) becomes
  $$\eta_G(\phi)\propto \Tr \left[(\widetilde X_1)^2
  +(\widetilde X_2)^2 +(\widetilde X_3)^2\right]\ .\eqno(3.10)$$
Since the three $\widetilde X_i$ are a basis for ${\cal SU}(2)$ (in
some representation), the quantity $\Tr (\widetilde X_i)^2$ is
the same for each $i$.  If we therefore pick, for example, $i=3$,
we may say
  $$\eta_G(\phi)\propto \Tr (\widetilde X_3)^2\ .\eqno(3.11)$$

We may notice a resemblance between (3.10) and the
expression for the second Casimir operator of the representation.
For an irreducible $d$-dimensional representation $\Lambda$, this
is a matrix
  $$\eqalign{C_2(\Lambda)&=(\Lambda_1)^2+(\Lambda_2)^2
  +(\Lambda_3)^2\cr &=J(J+1){\bf I}_d\ ,\cr}\eqno(3.12)$$
where the $\Lambda_i$ are the basis for ${\cal SU}(2)$ in this
representation, ${\bf I}_d$ is a $d\times d$ identity matrix, and
$J$ is the spin of the representation (related to $d$ by $J=(d-1)/2$).
Thus, if the representation $\widetilde X$ is irreducible,
the right hand side of (3.10) is just $\Tr C_2(\widetilde X)$; if it is
reducible, it will be a sum of $\Tr C_2(\Lambda^A)$ for each
irreducible factor $\Lambda^A$ in $\widetilde X$.  The trace
simply introduces a factor of $d$; hence, (3.10) becomes
  $$\eta_G(\phi)\propto \sum_{{\rm irreps}\ A}
  d_A(d_A-1)(d_A+1)\ .\eqno(3.13)$$
(Remember that the proportionality sign means that the winding
number equals the expression on the right hand side times constant
factors which do not depend on the map $\phi$.)
Notice that (3.13) allows us to determine what maps $\phi$ determine
the generator of $\p3(G)$; in principle, we can examine all
homomorphisms ${\cal SU}(2)\rightarrow {\cal G}$, and search
for the one for which (3.13) is the lowest.  In practice, it
is usually easy to find the generator.

Since the right hand sides of (3.11) and (3.13) are each proportional
to $\eta_G(\phi)$, they clearly must be proportional to each other.
It will be convenient to make this proportionality explicit for a
certain choice of conventions.  Thus, we will define
  $$\eqalign{\xi_G(\phi_*)&= -2\Tr(\widetilde X_3)^2\cr
  &={1\over 6}\sum_{{\rm irreps}\ A}
  d_A(d_A-1)(d_A+1)\ ,\cr}\eqno(3.14)$$
a number characterizing the algebra map $\phi_*:
{\cal SU}(2)\rightarrow {\cal G}$ (and the associated group
map $\phi$).
Note that the expression $\Tr(\widetilde X_3)^2$ is indeed
convention dependent; thus, (3.14) represents a statement about the
conventions we shall choose below.  Note also that the factor $1/6$
sets $\xi_G(\phi_*)=1$ when the representation of ${\cal SU}(2)$
induced by $\phi_*$ is two-dimensional and irreducible; this is the
lowest possible value for maps which define nontrivial
representations.  Finally,
it is important to distinguish between $\xi_G(\phi_*)$ and the
winding number $\eta_G(\phi)$; there is no guarantee that $\xi_G(g_*)$
will be unity for a map $g:\su2\rightarrow G$ which generates
$\p3(G)$, so we must normalize the winding number appropriately:
  $$\eta_G(\phi)={{\xi_G(\phi_*)}\over {\xi_G(g_*)}}\ .\eqno(3.15)$$

Therefore, in order to compute $p$ from (3.3), we need to choose
homomorphisms $f:\su2\rightarrow H$ and $g:\su2\rightarrow G$,
which are generators of $\p3$ in their respective groups
($\eta_H(f)=1$ and $\eta_G(g)=1$).\footnote{$^3$}{As $i:H
\rightarrow G$ is already a homomorphism, we need only to choose
$f$ to be a homomorphism $\su2\rightarrow H$ for $i\circ f$ to be
a homomorphism.}  We then find the associated maps (3.6)
between the Lie algebras, and take the ratio of (3.14) evaluated
for $i\circ f$ and $g$.  In Sec.~IV we shall go through this
procedure for various symmetry-breaking patterns.

{\bf B. Finding the Generators}

In order to apply the procedure just outlined, it is necessary to
find maps from $\su2$ to both $G$ and $H$ which generate the third
homotopy group.  As mentioned previously, it is sufficient to find
a homomorphism between the Lie algebra of $\su2$ and that of the
group; this will define the group homomorphism.  We now set about
finding the relevant maps for the cases of interest to us, namely
$\su{n}$, $\so{n}$ and the symplectic groups $\sp{n}$.

We begin by considering $\su2$.  Since $\su2$ is homeomorphic
to $S^3$, and $\p3(S^3)$ is generated by the identity map,
$\p3(\su2)$ is generated by the identity homomorphism.
Let us think about this fact on the Lie algebra level.
The usual convention in physics is to define a
Lie algebra element $T$ of $\su{n}$ to be
traceless and Hermitian, such that group elements are of the
form $g(T)=\exp(iT)$.  For convenience we shall use the notation
common in the mathematics literature, where the Lie algebra
consists of matrices $X$ which are traceless and skew-Hermitian
($X^\dagger =-X$), such that group elements are of the
form $g(X)=\exp(X)$.  Thus, we take the basis vectors $X_i$
of ${\cal SU}(2)$ in the defining representation to be
  $$X_1={1\over 2}\left(\matrix{0&i\cr i&0\cr}\right)\quad
  X_2={1\over 2}\left(\matrix{0&-1\cr 1&0\cr}\right)\quad
  X_3={1\over 2}\left(\matrix{i&0\cr 0&-i\cr}\right)\ ,\eqno(3.16)$$
with commutation relations
  $$[X_i,X_j]=\epsilon^{ijk}X_k\ .\eqno(3.17)$$
As this is an irreducible two-dimensional representation, the
(trivial) Lie algebra homomorphism $g_*:X_i\mapsto X_i$ has
  $$\xi_{\su2}(g_*)=1\ .\eqno(3.18)$$
(Note that, in the conventions we have chosen, both expressions
in (3.14) yield the same value for $\xi_{\su2}(g_*)$.)
This is the lowest possible value that $\xi$ can take for any
map, and hence the identity homomorphism $g:\su2\rightarrow\su2$
specified by $g_*$ is seen to be the generator of $\p3(\su2)$,
in accordance with expectation.

For all $\su{n}$ $n\geq 2$, we have $\p3(\su{n})=\Z$.  It is easy
to find a map which generates $\p3$, since there is a
natural map $g_*:{\cal SU}(2)\rightarrow {\cal SU}(n)$, given
by placing the $2\times 2$ matrices representing $X_i$
in the upper left hand corner of $n\times n$ matrices which are
zero elsewhere:
  $$g_*:X_i\mapsto \widetilde X_i \equiv
  \left(\matrix{\big(X_i\big)&&&\cr &0&&\cr
  &&0&\cr &&&0\cr}\right)\ .\eqno(3.19)$$
The Lie algebra elements $\widetilde X_i$, upon exponentiation,
define an $\su2$ subgroup of $\su{n}$ with an element of $\su2$
in the upper left hand corner, ones on the rest of the diagonal,
and zeroes elsewhere.  We therefore may extend (3.18) to
  $$\xi_{\su{n}}(g_*)=1\ ,\quad n\geq 2\ .\eqno(3.20)$$
Once again, as this is the minimum value $\xi$ may take,
the group homomorphism
$g:\su2\rightarrow \su{n}$ specified by (3.19) is a generator
of $\p3(\su{n})$.

The case of $\so{n}$ is somewhat more interesting.  $\so2$ is
topologically a circle, and thus $\p3(\so2)=0$.  In the case
of $\so3$, on the other hand, the Lie algebra is isomorphic to
that of $\su2$, and this isomorphism specifies a map $g:\su2
\rightarrow \so3$ which generates $\p3(\so3)$.  This map is the
familiar double cover of $\so3$ by $\su2$; indeed, we may think
of $\so3$ as $\su2/\Z_2$, in which case the exact homotopy
sequence
  $$\matrix{\p3(\Z_2)&\mapright{}&\p3(\su2)&\mapright{\beta}&
  \p3(\so3)&\mapright{}&\p2(\Z_2)\cr \Vert&&\Vert &&
  \Vert &&\Vert\cr 0&&\Z &&\Z &&0\cr}\eqno(3.21)$$
tells us immediately that the map $\beta$ is an isomorphism.
Thus, the generator of $\p3(\so3)$ is a map from the three-sphere
into $\so3$ which wraps twice around the group.  Let's see how
this appears at the Lie algebra level.  The Lie
algebra of $\so{n}$ is given by antisymmetric real $n\times n$
matrices.  For $n=3$, we may choose as a basis three matrices
$Y_i$, given by
  $${Y_1=\left(\matrix{0&-1&0\cr 1&0&0\cr 0&0&0\cr}\right)\quad
  Y_2=\left(\matrix{0&0&-1\cr 0&0&0\cr 1&0&0\cr}\right)\quad
  Y_3=\left(\matrix{0&0&0\cr 0&0&-1\cr 0&1&0\cr}\right)\ ,}
  \eqno(3.22)$$
which have commutation relations
  $$[Y_i,Y_j]=\epsilon^{ijk}Y_k\ .\eqno(3.23)$$
The isomorphism $g_*:{\cal SU}(2)\rightarrow{\cal SO}(3)$ is
therefore immediate:
  $$g_*:X_i\mapsto Y_i\ .\eqno(3.24)$$
This isomorphism establishes the matrices (3.22) as a representation
of ${\cal SU}(2)$; it is three-dimensional and irreducible.  Thus,
  $$\xi_{\so3}(g_*)=4\ .\eqno(3.25)$$
We can be certain that the group homomorphism defined by $g_*$
does generate $\p3(\so3)$, as there are no two-dimensional
representations of ${\cal SU}(2)$ by real antisymmetric $3\times 3$
matrices.  Therefore no other homomorphism ${\cal SU}(2)\rightarrow
{\cal SO}(3)$ will yield a lower value for $\xi$.

$\so4$ is something of a special case, in that it is not a simple
group.  The Lie algebra ${\cal SO}(4)$ is isomorphic to ${\cal SU}(2)
\oplus{\cal SU}(2)$, and $\so4$ itself is isomorphic to $(\su2\times
\su2)/\Z_2$.  This latter fact allows us to use the exact homotopy
sequence to find $\p3(\so4)$:
  $$\matrix{\p3(\Z_2)&\mapright{}&\p3(\su2\times \su2)&\mapright{\beta}&
  \p3(\so4)&\mapright{}&\p2(\Z_2)\cr \Vert&&\Vert &&
  \Vert &&\Vert\cr 0&&\Z\oplus\Z &&\Z\oplus\Z &&0\cr}\ .\eqno(3.26)$$
Thus, there are two generators of $\p3(\so4)$, each of which may be
thought of as inherited from a projection map $p:\su2\times\su2
\rightarrow \so4$.  These generators may be specified by exhibiting
the Lie algebra isomorphism $p_*:{\cal SU}(2)\oplus{\cal SU}(2)\rightarrow
{\cal SO}(4)$.  A basis for ${\cal SU}(2)\oplus{\cal SU}(2)$ is given
by $4\times 4$ matrices $X_i^a$, $X^b_i$, $i=1,2,3$, where the $X^a_i$
consist of the ${\cal SU}(2)$ generators $X_i$ from (3.16) in the upper
left $2\times 2$ block and zeroes elsewhere, while the $X^b_i$
have the $X_i$ in the lower right block and zeroes elsewhere.
As a basis for ${\cal SO}(4)$, we choose
  $$\eqalign{Z_1&={1\over 2}\left(\matrix{\hfill 0&-1&\hfill 0&\hfill 0\cr
  \hfill 1&\hfill 0&\hfill 0&\hfill 0\cr
  \hfill 0&\hfill 0&\hfill 0&\hfill 1\cr
  \hfill 0&\hfill 0&-1&\hfill 0\cr}\right)\quad
  Z_2={1\over 2}\left(\matrix{\hfill 0&\hfill 0&-1&\hfill 0\cr
  \hfill 0&\hfill 0&\hfill 0&-1\cr
  \hfill 1&\hfill 0&\hfill 0&\hfill 0\cr
  \hfill 0&\hfill 1&\hfill 0&\hfill 0\cr}\right)\quad
  Z_3={1\over 2}\left(\matrix{\hfill 0&\hfill 0&\hfill 0&-1\cr
  \hfill 0&\hfill 0&\hfill 1&\hfill 0\cr
  \hfill 0&-1&\hfill 0&\hfill 0\cr
  \hfill 1&\hfill 0&\hfill 0&\hfill 0\cr}\right)\cr
  Z_4&={1\over 2}\left(\matrix{\hfill 0&\hfill 1&\hfill 0&\hfill 0\cr
  -1&\hfill 0&\hfill 0&\hfill 0\cr
  \hfill 0&\hfill 0&\hfill 0&\hfill 1\cr
  \hfill 0&\hfill 0&-1&\hfill 0\cr}\right)\quad
  Z_5={1\over 2}\left(\matrix{\hfill 0&\hfill 0&\hfill 1&\hfill 0\cr
  \hfill 0&\hfill 0&\hfill 0&-1\cr
  -1&\hfill 0&\hfill 0&\hfill 0\cr
  \hfill 0&\hfill 1&\hfill 0&\hfill 0\cr}\right)\quad
  Z_6={1\over 2}\left(\matrix{\hfill 0&\hfill 0&\hfill 0&\hfill 1\cr
  \hfill 0&\hfill 0&\hfill 1&\hfill 0\cr
  \hfill 0&-1&\hfill 0&\hfill 0\cr
  -1&\hfill 0&\hfill 0&\hfill 0\cr}\right)\ .\cr}
  \eqno(3.27)$$
It is easy to check that the map $p_*:X_i^a\mapsto Z_i$, $X_i^b\mapsto
Z_{i+3}$ is an algebra isomorphism.  Thus, the submanifolds defined
by exponentiation of the two subalgebras ($Z_1$, $Z_2$, $Z_3$) and
($Z_4$, $Z_5$, $Z_6$) serve as the two generators of $\p3(\so4)$.
As representations of ${\cal SU}(2)$, these subalgebras are each
reducible into a sum of two two-dimensional representations.
We can evaluate (3.14) for the homomorphism $p_*$ restricted to
either the $X^a_i$ or the $X^b_i$, obtaining
  $$\xi_{\so4}[p_*(X^a_i)]=\xi_{\so4}[p_*(X^b_i)]=2\ .\eqno(3.28)$$
Notice that this number is lower than
the result (3.25) for $\so3$.  This is because the representation of
${\cal SU}(2)$ given by ($Z_1$, $Z_2$, $Z_3$) cannot fit inside
${\cal SO}(3)$; the sum of two irreducible two-dimensional
representations cannot be expressed as real $3\times 3$ antisymmetric
matrices.  With $4\times 4$ matrices there is no difficulty, and
we are able to find a representation for which $\xi$ is lower.

For $n>4$, the $\so{n}$ are all simple, and hence $\p3(\so{n})=\Z$.
We may choose the first six basis elements for ${\cal SO}(n)$ to
be $n\times n$ matrices with the matrices $Z_i$
from (3.27) in the upper left corner and zeroes elsewhere.  (We will
refer to the resulting matrices also as $Z_i$; no confusion
should arise.)  A homomorphism ${\cal SU}(2)\rightarrow
{\cal SO}(n)$ is given by
  $$g_*:X_i\mapsto Z_i\ ,\quad i=1,2,3\ .\eqno(3.29)$$
Just as in (3.28), this representation has $\xi_{\so{n}}(g_*)=2$,
which guarantees that the
group homomorphism $g$ defined by the algebra homomorphism
$g_*$ is a generator of $\p3(\so{n})$, as there is no two-dimensional
representation of ${\cal SU}(2)$ via real antisymmetric $n\times n$
matrices.  On the other hand, we could also consider the
homomorphism $g_*^\prime:X_i\mapsto Z_{i+3}$, $i=1,2,3$.  This
also defines a generator of $\p3(\so{n})$; however, unlike in
$\so4$, in $\so{n>4}$ these two generators are actually
homotopic to each other.  Therefore we may take either one as
the generator, and we shall generally choose (3.29).

Finally we consider the symplectic groups $\sp{2n}$, all
of which are simple.\footnote{$^4$}{We
are taking $\sp{2n}$ to be the symplectic group consisting of
$2n\times 2n$ matrices, with $n$ an integer.  Other conventions
denote this same group by $\sp{n}$.}  As basis elements for the
algebra ${\cal SP}(2n)$ we may choose
  $${\bf I}_2\otimes A^{(n)}_j\ ,\quad
  X_i\otimes S^{(n)}_k\ ,\eqno(3.30)$$
where ${\bf I}_2$ is a $2\times 2$ identity matrix, the $X_i$ are the
$2\times 2$ elements of ${\cal SU}(2)$ defined in (3.16), the $A^{(n)}_j$
($j=1,\ldots {1\over 2}n(n-1)$) are a basis for real antisymmetric
$n\times n$ matrices, and the $S^{(n)}_k$ ($k=1,\ldots {1\over
2}n(n+1)$) are a basis for real symmetric $n\times n$ matrices.
Note that all of these matrices are traceless and skew-Hermitian;
hence, ${\cal SP}(2n)$ is naturally a subalgebra of ${\cal SU}(2n)$.
Let $S^{(n)}_1$ be the $n\times n$ matrix with a $1$ in the upper
left corner and zeroes elsewhere.  Then there is a homomorphism
$g_*:{\cal SU}(2)\rightarrow{\cal SP}(2n)$ given by
  $$g_*:X_i\rightarrow X_i\otimes S^{(n)}_1\ .\eqno(3.31)$$
For this (two-dimensional) representation we have
  $$\xi_{\sp{2n}}(g_*)=1\ .\eqno(3.32)$$
As this is the lowest value possible, the group homomorphism
specified by (3.31) must be a generator of $\p3(\sp{2n})$.

\vskip .5cm
\noindent{\bf IV. Examples}

In this section we apply the general formulae just derived to
explicit computation of $\p3(G/H)$ for different choices of
$G$ and $H$.  We shall consider the cases $G=\su{n}$ and
$G=\so{n}$.  For these groups, Li [12] has calculated what
subgroup $H$ remains unbroken when a set of scalar fields
$\phi$ transforming under a specified representation of $G$
attains a specified vacuum expectation value.  In the
interest of completeness
we also compute $\p1(G/H)$ and $\p2(G/H)$.

It is important to note that a theory invariant under global
$\su{n}$ or $\so{n}$ transformations is sometimes invariant under
the larger symmetry groups $\u{n}$ or $\o{n}$, respectively.  The
existence of this extra degree of symmetry may affect the
corresponding vacuum manifold, if the extra symmetry is broken
by a vaccum expectation value.  In what follows we will assume
that the true symmetry really is $\su{n}$ or $\so{n}$; the
generalization is straightforward.  In any event, only the first
and second homotopy groups of the vacuum manifold can depend on this
distinction, and therefore the prediction of textures is unaltered.

{\bf A. $G=\su{n}$}

Consider a set of scalar fields $\phi^a$ ($a=1\ldots n$), which
transform in the vector representation of a global symmetry group
$\su{n}$:
  $$\phi^a\rightarrow \phi^{a^\prime}=U^{a^\prime}{}_a\phi^a ,
  \eqno(4.1)$$
where $U$ is an $n\times n$ matrix representing
an element of $\su{n}$.  Now imagine that $\phi^a$ attains a
vacuum expectation value
  $$\langle\phi^a\rangle=\rho\delta_{an}\ ,\eqno(4.2)$$
where $\rho$ is a constant parameter and $\delta$ is the Kronecker
delta.  The unbroken subgroup consists of those transformations
which leave (4.2) invariant; in this case, $H=\su{n-1}$,
as explained in Li [12].  We therefore
wish to compute $\p3(\su{n}/\su{n-1})$.  Our task
is facilitated by the fact that $\su{n}/\su{n-1}$ is homeomorphic
to $S^{2n-1}$ (Eq.~2.1).  We are only interested in $n\geq 2$
(since $\su0$ does not exist), and hence in spheres $S^q$ with
$q\geq 3$.  In this case $\p3$ follows immediately from (2.2):
  $$\p3(\su{n}/\su{n-1})=\cases{\Z & for $n=2$ ,\cr
  0 & for $n>2$ \ .\cr}\eqno(4.3)$$
We can derive this same result using the techniques of Sec.~III.  There
we found that the generator of $\p3(\su{n})$ could be taken to be
an $\su2$ subgroup in the upper left corner of the $\su{n}$ matrices.
Since the subgroup $H=\su{n-1}$ consists of $(n-1)\times (n-1)$
matrices in the upper left corner, it is clear that (for $n\geq 3$)
the map $g:\su2\rightarrow \su{n}$
generating $\p3(\su{n})$ is precisely the same as the
composition $i\circ f:\su2\rightarrow\su{n}$, where $f$ is the
generator of $\p3(\su{n-1})$ and $i:\su{n-1}\hookrightarrow \su{n}$ is the
inclusion of $H$ in $G$.  Therefore (3.3) becomes
  $$p={{\xi_{\su{n}}(i\circ f)}\over{\xi_{\su{n}}(g)}}=1
  \ .\eqno(4.4)$$
According to (3.2), $p$ serves to determine $\p3(G/H)=\Z_p$;
hence, $\p3(\su{n}/\su{n-1})=\Z_1=0$ when $n\geq 3$.
(``$\Z_1$,'' the integers
modulo division by $1$, is the trivial group.)  Thus (4.3) is
verified.

We next consider the case of $l$ distinct
$n$-vectors $\phi^a_j$ ($j=1\ldots l$),
all of which acquire linearly independent vacuum expectation
values.  We may take all of the nonzero components of $\langle
\phi^a_j\rangle$ to be $a=n-l+1,n-l,\ldots n$.
The original $\su{n}$ symmetry is broken to $H=\su{n-l}$.  Once
again, for $n-l\geq 2$, the maps $f$ (generating $\p3(H)$) and $g$
(generating $\p3(G)$) satisfy $i\circ f=g$, where $i$ is the inclusion
of $H$ in $G$.  Thus, (4.4) holds true for this case as well, and we
obtain
  $$\p3(\su{n}/\su{n-l})=0\ ,\qquad n-l\geq 2\ .
  \eqno(4.5)$$
The same result may be obtained by the factorization procedure of
the Appendix, by considering the map $i:\su{n-l}\rightarrow \su{n}$
as the composition $i_{n-1}\circ i_{n-2}\circ\ldots\circ i_{n-l}$,
where $i_m:\su{m}\hookrightarrow\su{m+1}$ is the usual inclusion
into the upper left corner.  Then the map between homotopy groups
$\alpha:\p3(\su{n-l})\rightarrow\p3(\su{n})$ factors similarly;
however, since $\p3(\su{m+1}/\su{m})=0$ (for $m\geq 2$),
each $\alpha_m:\p3(\su{m})\rightarrow\p3(\su{m+1})$
must be an isomorphism (by Eq.~(2.6)).  Thus $\alpha$ itself is
an isomorphism, and (4.5) is recovered.

We now consider scalar fields transforming in the symmetric
second-rank tensor representation,\footnote{$^5$}{Fields
transforming in the symmetric or antisymmetric tensor
representations may be invariant under an extra degree of symmetry,
making the full symmetry group $U(n)$ rather than $\su{n}$.
This may affect $\p1$ of the vacuum manifold, but will not
affect $\p3$, and we will not consider it in this paper.}
  $$\phi_{a^\prime b^\prime}=U^a{}_{a^\prime}
  U^{b}{}_{b^\prime}\phi_{ab}\ ,\eqno(4.6)$$
where $\phi_{ab}=\phi_{ba}$.  Such a fields may attain
a vacuum expectation value
  $$\langle\phi_{ab}\rangle=\rho\delta_{an}
  \delta_{bn}\ .\eqno(4.7)$$
This serves to break $\su{n}$ to $\su{n-1}$, and the analysis is
identical to the vector case, culminating in (4.3).  However,
a different choice of potential $V(\phi_{ab})$ can lead to a
vacuum expectation value
  $$\langle\phi_{ab}\rangle=\rho\delta_{ab}\ ,\eqno(4.8)$$
with $\rho$ once again a constant [12].  In this case the
unbroken symmetry group is $H=\so{n}$; note that $\so{n}$ is
naturally a subgroup of $\su{n}$, as all real orthogonal matrices
are automatically unitary.  The appearance of monopoles in this
model was studied in Ref.~[13].  We can proceed to compute $\p3(
\su{n}/\so{n})$ on a case-by-case basis.  For $n=2$, $\so2$ is
abelian, and hence $\p3(\su2/\so2)=\p3(\su2)=\Z$.  For $n=3$,
we recall from Sec.~II.B that the generator $f$ of $\p3(\so3)$
is a double cover, with $\xi_{\so3}(f_*)=4$ (3.25).  Since the
inclusion $i:\so3\hookrightarrow\su3$ is trivial, we will also
have $\xi_{\su3}[(i\circ f)_*]=4$.  Meanwhile, the generator $g$
of $\p3(\su3)$ has $\xi_{\su3}(g_*)=1$.  It then follows from
(3.14) that the winding number of $i\circ f$ in $G$ is 4, from
which we obtain $\p3(\su3/\so3)=\Z_4$.  The case $n\geq 5$ is
equally straightforward; the only difference is that in this case
$\xi_{\so{n}}(f_*)=2$, leading to $\p3(\su{n}/\so{n})=\Z_2$,
$n\geq 5$.  The only subtle case is $n=4$, due to the fact that
$\so4$ is not simple, and $\p3(\so4)=\Z\oplus\Z$.  Nevertheless,
the logic leading up to (2.6) is still valid, so we need only
to find the image of $\alpha:\p3(\so4)\rightarrow\p3(\su4)$.
As this is a homomorphism from $\Z\oplus \Z$ to $\Z$, it must
be of the form $\alpha:(a,b)\mapsto q_aa+q_bb$ for some integers
$q_a$, $q_b$.  We have already discovered in Sec.~III.B that the
generators of $\p3(\so4)$ (corresponding to $(a,b)=(1,0)$ and
$(0,1)$) may be obtained by exponentiating the subalgebras
($Z_1$, $Z_2$, $Z_3$) and ($Z_4$, $Z_5$, $Z_6$) as specified in
(3.27), and furthermore that (3.14) evaluated on either generator
gives $\xi_{\so4}(f_*)=2$, where $f_*$ is taken to be the map
from $\su2$ to either generator.  Thus, we derive $q_a=q_b=2$.
The image of $\alpha$ is therefore the even numbers; from (2.6)
we find $\p3(\su4/\so4)=\Z_2$.  Taken together, we have found
  $$\p3(\su{n}/\so{n})=\cases{\Z\ ,\quad n=2\cr
  \Z_4\ ,\quad n=3\cr \Z_2\ ,\quad n\geq 4\ .\cr}\eqno(4.9)$$
We therefore have the interesting situation where $\p3$ can be a
finite group, and two textures with positive winding number can
mutually unwind rather than collapsing.  We comment briefly on
this later.

We turn next to the antisymmetric second-rank tensor representation,
  $$\phi_{a^\prime b^\prime}=U^a{}_{a^\prime}
  U^{b}{}_{b^\prime}\phi_{ab}\ ,\eqno(4.10)$$
with $\phi_{ab}=-\phi_{ba}$.  There are two types of vacuum
expectation value which can be attained.  The first, for which
  $$\langle\phi_{ab}\rangle=\rho\left(\matrix{
  \left(\matrix{0&1\cr -1&0\cr}\right)&&&&0&\cr
  &0&&&&\cr &&.&&&\cr &&&.&&\cr 0&&&&.&\cr
  &&&&&0\cr}\right)\ ,\eqno(4.11)$$
breaks $\su{n}$ to $H=\su{n-2}\times\su2$; thinking of $G/H$
as $[\su{n}/\su{n-l}]/\su2$, it follows from (4.5) that
$\p3(G/H)=0$.  The other alternative is
  $$\langle\phi_{ab}\rangle=\rho\left(\matrix{
  \left(\matrix{0&1\cr -1&0\cr}\right)&&&&&0\cr
  &\left(\matrix{0&1\cr -1&0\cr}\right)&&&&\cr
  &&.&&&\cr &&&.&&\cr &&&&.&\cr
  0&&&&&\left(\matrix{0&1\cr -1&0\cr}\right)\cr}\right)\eqno(4.12)$$
when $n=2k$ is even, and
  $$\langle\phi_{ab}\rangle=\rho\left(\matrix{
  \left(\matrix{0&1\cr -1&0\cr}\right)&&&&&0&\cr
  &\left(\matrix{0&1\cr -1&0\cr}\right)&&&&&\cr
  &&.&&&&\cr &&&.&&&\cr &&&&.&&\cr
  0&&&&&\left(\matrix{0&1\cr -1&0\cr}\right)&\cr
  &&&&&&\cr
  &&&&&&0\cr}\right)\eqno(4.13)$$
when $n=2k+1$ is odd.  This pattern breaks $\su{2k}$ or $\su{2k+1}$
to $H=\sp{2k}$.  As with the orthogonal groups, the symplectic
matrices are automatically unitary, so the inclusion $i:\sp{2k}
\hookrightarrow\su{2k[+1]}$ is trivial.  Since we found in Sec.~III
that the generator $f$ of $\p3(\sp{2k})$ satisfied $\xi_{\sp{2k}}(f_*)
=1$, we must also have $\xi_{\sp{2k}}[(i\circ f)_*]=1$.  Hence the
winding number of $i\circ f$ is one, which implies
  $$\p3(\su{2k[+1]}/\sp{2k})=0\ ,\quad k\geq 1\ .\eqno(4.14)$$
This symmetry breaking pattern therefore does not lead to texture.

The final representation of $\su{n}$ we consider is the adjoint
representation, which transforms as
  $$\phi^{a^\prime}{}_{b^\prime}\rightarrow U^{\dagger a^\prime}{}_a
  U^{b}{}_{b^\prime}\phi^{a}{}_b\ ,\eqno(4.15)$$
where $\phi^{a}{}_{b}$ is an $n\times n$ skew-Hermitian
matrix, and the dagger denotes Hermitian conjugation.
The scalar fields $\phi^a{}_b$ can
attain a vacuum expectation value of the form
  $$\langle\phi^a{}_b\rangle=\left(\matrix{\rho_1&&&&&&\cr
  &\rho_1&&&&0&\cr &&.&&&&\cr &&&.&&&\cr &&&&.&&\cr
  &0&&&&\rho_2&\cr &&&&&&\rho_2\cr}\right)\ ,\eqno(4.16)$$
where $\rho_1$ is a constant appearing $l$ times, and
$\rho_2$ is a constant appearing $n-l$ times.  (By convention we
can choose $n-l\geq l$.)  The result is to
break $\su{n}$ to $H={\rm S}[\u{n-l}\times\u{l}]=
[\su{n-l}\times\su{l}\times\u1]/\Z_{2(n-l)}$, leading to
vacuum manifolds known as Grassmann spaces.  This symmetry
breaking pattern is well-known from grand unified theories, where
$\su5$ is often said to break to $\su3\times\su2\times\u1$.
While this is true at the Lie algebra level, the unbroken subgroup
is actually ${\rm S}[\u{n-l}\times\u{l}]$, consisting of matrices
of the form
  $$M=\left(\matrix{A & 0_{l\times(n-l)}\cr
  0_{(n-l)\times l} & B\cr}\right)\ ,\quad \det M=1
  \ ,\eqno(4.17)$$
where $A\in\u{n-l}$, $B\in\u{l}$, and $0_{l\times(n-l)}$ is an
${l\times(n-l)}$ zero matrix.  When $n-l=l=1$, we have
${\rm S}[\u{1}\times\u{1}]=\u1$, and hence $\p3({\rm S}[\u{1}\times
\u{1}]=0$.  For $n-l>l=1$, we have
$\p3({\rm S}[\u{n-1}\times\u{1}])=\Z$, and for $n-l\geq l\geq 2$
we have $\p3({\rm S}[\u{n-l}\times\u{l}])=\Z\oplus\Z$.  In either
of the latter cases, one generator of $\p3$ is given by a subgroup
consisting of matrices with an $\su2$ matrix in the upper left
corner, ones on the remaining diagonal, and zeroes elsewhere.
As this subgroup also generates $\p3(\su{n})$, the map
$\alpha:\p3({\rm S}[\u{n-l}\times\u{l}])\rightarrow \p3(\su{n})$
will be onto.  Hence, from (2.6), we have
  $$\p3[\su{n}/{\rm S}[\u{n-l}\times\u{l}]]=\cases{
  \Z\ , & for $ n-l=l=1$ \cr
  0\ , & for $n\geq 3,\ n-l\geq l\geq 1$\ .\cr}\eqno(4.18)$$

{\bf B. $G=\so{n}$}

Since the orthogonal groups may be thought of as unitary groups over
the real numbers rather than the complex numbers, it should not be
surprising that much of the structure uncovered for the case $G=\su{n}$
repeats itself when $G=\so{n}$.  For example, we consider a set of $l$
distinct $n$-vectors $\phi^a_j$ ($j=1\ldots l$), all of which
transform as
  $$\phi^a_j\rightarrow \phi^{a^\prime}_j=O^{a^\prime}{}_a\phi^a_j ,
  \eqno(4.19)$$
where $O$ is an $n\times n$ matrix representing
an element of $\so{n}$.  If each vector attains a linearly independent
vacuum expectation value,  the symmetry will be spontaneously broken to
$\so{n-l}$.  In the case of $\su{n}$ the resulting vacuum manifolds
had trivial third homotopy groups (for $n-l\geq 2$), because the
generator of $\p3(G)$ could be thought of as the generator of $\p3(H)$
included in $G$.  The same will occur for $\so{n}$, but only when $n$
and $n-l$ are sufficiently large.  Consider the case $n> n-l
\geq 4$.  As usual, the inclusion $\so{n-l}\hookrightarrow\so{n}$
places an element of $\so{n-l}$ into the upper left corner of the
$\so{n}$ matrix.  We have discovered that the generators of $\p3(\so{n})$
and $\p3(\so{n-l})$ are the same, and are specified by exponentiation of
the Lie subalgebra with basis elements $Z_1$, $Z_2$, $Z_3$ from (3.27).
(This is not precisely true for $n-l=4$, since $\p3(\so4)$ has two
generators; nevertheless, the important fact is that one of the
generators of $\p3(\so4)$ also generates $\p3(\so{n})$.)  Thus, just
as in (4.4) for the case of $\su{n}$, we obtain
  $$p={{\xi_{\so{n}}(i\circ f)}\over{\xi_{\so{n}}(g)}}=
  {2\over 2}=1\ ,
  \quad n>n-l\geq 4 \ ,\eqno(4.20)$$
leading via (3.2) to
  $$\p3(\so{n}/\so{n-l})=0\ ,\quad n>n-l\geq 4 \ .\eqno(4.21)$$
The lower-dimensional cases must be handled individually.  For $n\geq5$
and $n-l=3$, the procedure is the same, but the generator of $\p3(\so3)$
is specified by the Lie algebra (3.22).  Hence $\xi_{\so{n}}(i\circ f)
=4$, and
  $$\p3(\so{n}/\so{3})=\Z_2\ ,\quad n\geq 5 \ .\eqno(4.22)$$
Meanwhile, when $n=4$ and $n-l=3$, we use $\so4/\so3\sim S^3$ to obtain
  $$\p3(\so4/\so{3})=\Z\ .\eqno(4.23)$$
The only remaining possibility is $n-l=2$; as $\so2$ is abelian we have
  $$\p3(\so{n}/\so2)=\p3(\so{n})\ .\eqno(4.24)$$

A set of fields $\phi_{ab}=\phi_{ba}$ transforming in the symmetric
second-rank tensor representation of $\so{n}$ can attain a vacuum
expectation value for which $\phi_{ab}$ is diagonal; details can be
found in [12].  The result is to break $\so{n}$ to a subgroup
$H={\rm S}[\o{n-l}\times\o{l}]=\so{n-l}\times\so{l}\times\Z_2$,
where we can always choose $n-l\geq l$.
As in the case $H=\so{n-l}$, the analysis is straightforward for
sufficiently large $n$, $n-l$; using (2.9), we will ignore the $\Z_2$
factor.  Specifically, for $n>n-l\geq 4$,
we have the exact sequence
  $$\matrix{\p3(\so{n}/\so{n-l})&\mapright{}&
  \p3(\so{n}/(\so{n-l}\times\so{l}))&\mapright{}
  &\p2(\su{l})\cr \Vert &&
  &&\Vert\cr 0&& &&0\cr}\ ,\eqno(4.25)$$
which implies
  $$\p3[\so{n}/(\so{n-l}\times\so{l})]=0\ ,
  \qquad n>n-l\geq 4\ .\eqno(4.26)$$
There are only three cases for which $n-l\leq 3$:
$n=6$, $l=3$; $n=5$, $l=2$; and $n=4$, $l=2$.  In the last of these,
$H=\so2\times\so2$ is abelian, and we have
  $$\p3[\so4/(\so2\times\so2)]=\p3(\so4)=\Z\oplus\Z\ .\eqno(4.27)$$
Similarly, for $G=\so5$, $H=\so3\times\so2$, the abelian factor is
irrelevant, and
  $$\p3[\so5/(\so3\times\so2)]=\p3(\so5/\so3)=\Z_2\ .\eqno(4.28)$$
Lastly, the case $G=\so6$, $H=\so3\times\so3$ bears a close
resemblance to $G=\su4$, $H=\so4$.  From (2.6), we need only to
find the image of
  $$\matrix{\p3(\so3\times\so3)&\mapright{\alpha}&
  \p3(\so6)\cr \Vert &&
  \Vert\cr \Z\oplus\Z &&\Z \cr}\ .\eqno(4.29)$$
The same reasoning used in the computation of $\p3(\su4/\so4)$ tells
us that the image of $\alpha$ is the even numbers, which implies
  $$\p3[\so6/(\so3\times\so3)]=\Z_2\ .\eqno(4.30)$$

The last case to consider is that of the antisymmetric second-rank
tensor representation $\phi_{ab}=-\phi_{ba}$.  As in the case of
$\su{n}$, the scalar fields may attain a vacuum expectation value
of the form (4.11), breaking $\so{n}$ to $\so{n-2}\times \so2$; we have
computed $\p3(\so{n}/(\so{n-2}\times\so2))$ above.  The fields may also
attain a vacuum expectation value of the form (4.12) if $n=2k$ is even,
and (4.13) if $n=2k+1$ is odd ($k\geq2$).  In either case, the unbroken
subgroup is $H=\u{k}$.  Let us consider the Lie algebra map
$i_*:{\cal U}(k)\rightarrow{\cal SO}(2k)$ corresponding to the inclusion
$i$ of $\u{k}$ into $\so{2k}$.  It takes a complex $k\times k$
skew-Hermitian matrix $X$, which we write $X_R+iX_I$, to a real
antisymmetric $2k\times 2k$ matrix of the form
  $$i_*:X_r+iX_I\mapsto {\widetilde X}\equiv
  \left(\matrix{1&0\cr 0&1\cr}\right)
  \otimes X_R +\left(\matrix{\ 0&1\cr -1&0\cr}\right)\otimes X_I
  \ .\eqno(4.31)$$
For $n=2k+1$ the map is the same, with an extra column and row of zeroes
added to $\widetilde X$.  Since the map $f_*:{\cal U}(2)\rightarrow
{\cal U}(k)$ is given by (3.19), it is straightforward to compute
that
  $$\xi_{\so{2k[+1]}}(i_*\circ f_*)=2\ .\eqno(4.32)$$
As the generator $g$ of $\p3(\so{n})$ satisfies $\xi_{\so{n}}(g_*)=2$
for all $n\geq 5$, we have
  $$\p3(\so{2k[+1]}/\u{k})=0\ ,\quad 2k[+1]\geq 5\ .\eqno(4.33)$$
The only other case of interest is $\so4/\u2$, for which precisely
the same logic holds, with the exception that $\p3(\so4)=\Z\oplus\Z$;
therefore, the image of $\alpha:\p3(\u2)\rightarrow\p3(\so4)$ is one
of the $\Z$ factors, and
  $$\p3(\so4/\u2)=\Z\ .\eqno(4.34)$$
This completes our computation of $\p3(G/H)$.

{\bf C. Other homotopy groups}

The various symmetry breaking patterns we have studied may lead not
only to textures, but also to strings, monopoles and domain walls.
Here we calculate the lower homotopy groups $\p1$ and
$\p2$ of the vacuum manifolds $G/H$ considered above, to determine
whether these theories predict strings or monopoles.  (We do not study
$\p0$, which governs the appearance of domain walls, as all of the
groups we consider are connected, and $\p0(G/H)$ is always the trivial
group.)  For the most part, these calculations require less effort than
the computation of $\p3$.

For all Lie groups $G$, $\p2(G)=0$; furthermore, for most of the groups
we consider (those without $Z_n$ factors), $\p0(G)=0$.
We therefore have the following exact homotopy sequence:
  $$0\ \mapright{\beta}\ \p{2}(G/H)\
  \mapright{\gamma}\ \p{1}(H)\ \mapright{\bar\alpha}
  \ \p{1}(G)\ \mapright{\bar\beta}\ \p{1}(G/H)
  \ \mapright{\bar\gamma}\ 0\ . \eqno(4.35)$$
For $G=\su{n}$, we have $\p1(G)=0$, which leads immediately to
  $$\p2(\su{n}/H)=\p1(H)\ ,\eqno(4.36)$$
and
  $$\p1(\su{n}/H)=0\ ,\eqno(4.37)$$
for any choice of $H$.

The case $G=\so{n}$, $n\geq3$, requires more effort.  Let's begin
with $H=\so{n-l}$, for $n-l\geq 3$.  We need to examine the
exact sequence (4.35) with $\p1(H)=\p1(G)=\Z_2$.  However, despite
the fact that $\p1(H)$ and $\p1(G)$ are isomorphic as groups,
it does not necessarily follow that the map $\bar\alpha$ is
itself an isomorphism.  Therefore this
sequence by itself is insufficient to compute the unknown
homotopy groups; we must examine the map $\bar\alpha$ in more
detail.  In a manner analogous to that discussed below (4.5)
for the case of $\su{n}$, we may factor $\bar\alpha$ into
$\bar\alpha_{n-1}\circ\bar\alpha_{n-2}\circ\ldots \circ
\bar\alpha_{n-l}$, where $\bar\alpha_m:\p1(\so{m})\rightarrow
\p1(\so{m+1})$. We therefore consider the sequence (4.35) with
$G=\so4$, $H=\so3$.
The quotient space is $\so4/\so3\sim S^3$, and
we know that $\p2(S^3)=\p1(S^3)=0$.  Thus, (4.35) guarantees
that $\bar\alpha_3 :\p1(\so3)\rightarrow\p1(\so4)$ is an
isomorphism.  Similar arguments suffice to show that, for any
$m\geq 3$, $\bar\alpha_m :\p1(\so{m})\rightarrow\p1(\so{m+1})$ is an
isomorphism, and thus that $\bar\alpha:\p1(\so{n-l})\rightarrow
\p1(\so{n})$ is an isomorphism for all $n>n-l\geq3$.
Thus, the image of $\bar\alpha$ is all of $\p1(\so{n})=\Z_2$.
Exactness of (4.35) implies that the kernel of $\bar\beta:
\Z_2\rightarrow\p1(\so{n}/\so{n-l})$ is all of $\Z_2$,
and that the image of $\bar\beta$ is the kernel of $\bar\gamma$,
namely all of $\p1(\so{n}/\so{n-l})$.  Together these imply
that
  $$\p1(\so{n}/\so{n-l})=0\ ,\quad n\geq n-l\geq 3\ .\eqno(4.38)$$
Similar reasoning leads to
  $$\p2(\so{n}/\so{n-l})=0\ ,\quad n\geq n-l\geq 3\ .\eqno(4.39)$$
Furthermore, we can repeat the procedure with $H=\so2$, the
only modification being $\p1(H)=\Z$. The results in this case
are
  $$\p1(\so{n}/\so2)=0\ ,\quad n\geq 3\ ,\eqno(4.40)$$
and
  $$\p2(\so{n}/\so2)=\Z\ ,\quad n\geq 3\ .\eqno(4.41)$$

For $G/H=\so{n}/(\so{n-l}\times\so{l}\times\Z_2)$ we have $\p0(H)=\Z_2$,
and hence (4.35) does not apply.  Instead we may proceed in stages,
first considering $(\so{n}/\so{n-l})/\so{l}$ and then
$[\so{n}/(\so{n-l}\times\so{l})]/\Z_2$.  Without presenting the
relevant details, it is straightforward to find that the $\Z_2$
factor renders the vacuum manifold non-simply-connected:
  $$\p1(\so{n}/(\so{n-l}\times\so{l}\times\Z_2))=\Z_2\ ,\eqno(4.42)$$
while the second homotopy group is more complicated:
  $$\p2(\so{n}/(\so{n-l}\times\so{l}\times\Z_2))= \cases{
  \Z\ , & for $ n=3\ ,\ l=1$ \cr
  0\ , & for $ n\geq 4\ ,\ l=1$\cr
  \Z\oplus\Z\ , & for $n=4\ ,\ l=2$\cr
  \Z\ , & for $n\geq 5\ ,\ l=2$\cr
  \Z_2\ ,& for $n\geq 6\ ,\ l\geq 3$\ .\cr} \eqno(4.43)$$

Finally, we can consider $G=\so{2k[+1]}$ breaking down to
$H=\u{k}$.  In this case the exact sequence (4.35)
by itself is insufficient;
however, it is easy to check that the map $\bar\alpha$ is onto,
which then leads to
  $$\p1(\so{2k[+1]}/\u{k})=0 \ ,\eqno(4.44)$$
and
  $$\p2(\so{2k[+1]}/\u{k})=\cases{
  \Z\ , & for $ 2k[+1]\geq 3$ \cr
  0\ , & for $2k[+1]=2$\ .\cr} \eqno(4.45)$$

We have tabulated the results of all of our homotopy calculations
in Table Two.

\vskip .5cm

\noindent{\bf V. Discussion}

We have studied the topology of the vacuum manifolds $G/H$ resulting
from the spontaneous breakdown of a global symmetry $G$ to a subgroup
$H$.  The homotopy groups $\p{q}(G/H)$ are related to the existence
of field configurations of potential significance to cosmology:
topological defects for $q=0,1,2$, and textures for $q=3$.  Although
it is generally straightforward to calculate $\p{q}(G/H)$ for $q\leq
2$, the case $q=3$ relevant to textures is more difficult.  We have
therefore studied a number of different choices for $G/H$, and
computed the relevant homotopy groups.

Although the results of this paper allow us to predict whether any
given spontaneously broken global symmetry will lead to texture,
there are a number of questions remaining to be answered about the
cosmological effects of those textures which result.  For example,
one may ask what the likelihood is that a specified field
configuration will collapse and unwind, and how many such unwindings
per horizon volume are predicted by the Kibble mechanism.  In the
same vein, it would be of interest to know the characteristics of
collapse; {\it e.g.}, whether the field approaches a spherically symmetric
configuration, rather than a pancake or spindle configuration.  All
of these issues may in principle depend on the geometry of the
vacuum manifold $G/H$ under consideration.  Hence, both
analytic and numerical studies of the evolution of textures resulting
from different choices of $G$ and $H$ would be of interest.
(Such studies have been performed in the case of $G/H=S^3$ [14]
and $G/H=S^2$ [15].)

While we will not attempt to answer any of these questions in this
paper, we would like to briefly discuss the construction of field
configurations representing individual textures, which would be
appropriate initial conditions for simulations.  We imagine therefore
that we have a set of scalar fields $\Phi$ which transform under some
representation of the global symmetry group $G$.  We will write the
action of an element $\mu\in G$ on $\Phi$ as $\mu\Phi$, although the
explicit matrix form may be more complicated.  Our goal is to
specify a field configuration $\Phi(x)$ which is in the vacuum
manifold (or, equivalently, in $G/H$) and has a given winding number.

The action of $G$ on the vacuum manifold is transitive --- a fixed
element is taken to any other element by the action of the group.
Therefore, every field value $\Phi$ in the vacuum manifold
is of the form $\mu\Phi_0$, where we have specified a fiduciary field
value $\Phi_0$ in the vacuum manifold.  Hence, our sought-after
field configuration $\Phi(x)$ can be written
  $$\Phi(x)=\mu(x)\Phi_0\ ,\eqno(5.1)$$
where $\mu(x)$ is a map from space (at fixed time) to the symmetry group
$G$.  (Such a description may be highly redundant, as two different
group elements $\mu_1$ and $\mu_2$ may satisfy $\mu_1\Phi_0=\mu_2\Phi_0$;
however, this redundancy is not a concern in the construction of a
field configuration.)  We are only considering maps $\mu(x)$ which go to a
constant element of $G$ at spatial infinity; that is, $\mu(x)$
represents a map $S^3\rightarrow G$, and hence an element of $\p3(G)$.
Acting $\mu(x)$ on $\Phi_0$ therefore produces a map $S^3\rightarrow
G/H$, and hence an element of $\p3(G/H)$; indeed, we have just
exhibited the map $\beta:\p3(G)\rightarrow\p3(G/H)$ found in the exact
homotopy sequence.

Since we would like $\Phi(x)$ to represent a nonzero element of
$\p3(G/H)$, we are interested in elements of $\p3(G)$ which are not in
the kernel of $\beta$.  This is straightforward, given the results of
the previous sections, which allow us to construct the map $\beta$
explicitly by taking advantage of the fact that the kernel of $\beta$
was equal to the image of $\alpha:\p3(H)\rightarrow\p3(G)$.
For example, consider the case $G=\su3$, $H=\so3$, for
which $\p3(G)=\Z$ and $\p3(G/H)=\Z_4$, and $\beta$ is ``mod 4.''
Thus, elements of $\p3(\su3)$ with winding numbers 1, 2, or 3 will
be taken by $\beta$ to maps with the same winding number in
$\p3(\su3/\so3)$, while a map with winding number $4$ will be taken to
a map with winding number zero, and so on.  To specify a nonzero
element of $\p3(\su3/\so3)$, we therefore need only to find a map
$\mu(x)$ representing $\p3(\su3)$ with winding number one (for example).
This is also straightforward, as we have specified the generators of
$\p3(G)$ in Sec.~III.B, in terms of group homomorphisms
$\su2\rightarrow G$.  Thus, we specify a map $\tilde\mu(x)$ from $S^3$
(representing space) to $\su2$ which generates $\p3(\su2)$, and a map
$g:\su2\rightarrow G$ which generates $\p3(G)$; the field configuration
$\Phi=\mu(x)\Phi$, where $\mu(x)=g\circ\tilde\mu(x)$, will then have
winding number one (as long as $\beta$ takes a generator of $\p3(G)$
to a generator of $\p3(G/H)$).

Let us illustrate this procedure for $G=\su3$, $H=\so3$.  In a Cartesian
coordinate system $(x,y,z)$, an example of a map from
space to $\su2$ which covers
the group once (and hence generates $\p3(\su2)$) is given by
  $$\tilde\mu(x)={\bf I}\cos\chi(r) -i\vec\sigma\cdot{\hat x}
  \sin\chi(r)\ ,\eqno(5.2)$$
where ${\bf I}$ is a $2\times 2$ identity matrix, the $\sigma_i$
are the Pauli matrices (related to the Lie algebra elements $X_i$
of (3.16) by $\sigma_i=-2iX_i$), $r=\sqrt{x^2+y^2+z^2}$, ${\hat x_i}
=x_i/r$, and $\chi(r)$ is a function with boundary conditions
$\chi(0)=0$ and $\chi(\infty)=\pi$.  In polar coordinates $(r,\theta,
\phi)$ this becomes
  $$\tilde\mu(x)=\left(\matrix{\cos\chi-i\cos\theta\sin\chi &
  e^{i\phi}\sin\theta\sin\chi\cr -e^{-i\phi}\sin\theta\sin\chi &
  \cos\chi+i\cos\theta\sin\chi\cr}\right)\ .\eqno(5.3)$$
As usual, we will pick the generator of $\p3(\su3)$ to be represented
by the inclusion $g:\su2\rightarrow\su3$ in the upper left corner.
The scalar fields which break $\su3$ to $\so3$ lie in the symmetric
second-rank tensor representation of $\su3$, and attain a vacuum
expectation value of the form $\langle\phi_{ab}\rangle=\rho\delta_{ab}$,
with $\rho$ an arbitrary constant.  It is natural to choose as our
fiducial value of the field $\Phi_0=\langle\phi_{ab}\rangle$; then,
using the transformation law (4.6), the field configuration becomes
  $$\phi_{ab}(x)=\rho U^c{}_a(x)U^d{}_b(x)\delta_{cd}\ ,\eqno(5.4)$$
where $U^c{}_a(x)$ is an $\su3$ matrix representing $\mu(x)
=g\circ\tilde\mu(x)$.  Using (5.3), we therefore have
  $$\phi_{ab}(x)=\rho\left(\matrix{1-2ic_\theta c_\chi s_\chi
  -2s_\chi^2+(1+e^{-2i\phi})s_\theta^2s_\chi^2 &
  2i(s_\phi c_\chi-c_\phi c_\theta s_\chi)s_\theta s_\chi & \ 0\ \cr
  2i(s_\phi c_\chi-c_\phi c_\theta s_\chi)s_\theta s_\chi &
  1+2ic_\theta c_\chi s_\chi -2s_\chi^2+(1+e^{2i\phi})
  s_\theta^2s_\chi^2 & \ 0\ \cr
  0&0&\ 1\ \cr}\right)\ ,\eqno(5.5)$$
where $c_\chi\equiv\cos\chi$ and so on.  This configuration
represents a texture of winding number one, the evolution
of which could be studied numerically.

{}From this point, it is easy to construct additional configurations
with the same winding number, simply by choosing different maps
$\mu:S^3\rightarrow G$ which generate $\p3(G)$.  For example, for
any submanifold $\Sigma$ of $G$, conjugation by a fixed element $m$
defines a new submanifold $m\Sigma m^{-1}$ homotopic to $\Sigma$;
hence, if $\mu(x)$ generates $\p3(G)$, so will $\mu^\prime(x)=
m\mu(x)m^{-1}$.  Furthermore, given two maps $\mu_1(x)$ and $\mu_2(x)$,
the winding number in $\p3(G)$ obeys
  $$\eta_G(\mu_1\mu_2)=\eta_G(\mu_1)+\eta_G(\mu_2)\ .\eqno(5.6)$$
Therefore, field configurations with higher winding
numbers are readily constructed (if they exist at all).  These could
be single textures with winding number greater than one, or two
nearby textures.

Although it is still unclear whether topological properties of
spontaneously broken symmetries play a role in the formation of
large-scale structure in the universe, the lack of a single compelling
model of structure formation encourages further study of many
different models.  The calculations performed in this paper provide
a starting point for the study of a number of models beyond those
considered to date; further work should enable us to determine
the relationship of these theories to the observed universe.

\noindent{\bf Acknowledgements}

We would like to thank Larry Widrow for encouragement,
Alan Guth and Andrew Sornborger for useful comments, and especially
Sidney Coleman for invaluable help with our $\pi$'s.  This work
was supported in part by NASA under Grants no. NAGW-931 and
NGT-50850, by the National Science Foundation under grants
AST/90-05038 and PHY/92-06867, and by the U.S. Department
of Energy (D.O.E.) under contract no. DE-AC02-76ER03069.

\vskip .6cm

\noindent{\bf Appendix:  Factorization}

Whenever we have two spaces $X$ and $Y$ and a map $\phi:X\rightarrow
Y$, there is an induced map in homotopy $\Phi:\p{q}(X)\rightarrow\p{q}(Y)$.
In calculating the action of $\Phi$,
we are often aided by the existence of a third space $Z$ in between $X$
and $Y$, in the sense that there are maps $\psi_1:X\rightarrow Z$ and
$\psi_2:Z\rightarrow Y$ (which induce maps $\Psi_1$ and $\Psi_2$ in
homotopy) such that $\phi=\psi_2\circ\psi_1$.  In that case,
we can factor the map $\Phi$ into
$\Psi_2\circ\Psi_1$.  In other words, if
the diagram between topological spaces
  $$\matrix{X&\mapright{\psi_1}&Z\cr
    &_{\phi}\!\searrow\ &\mapdown{\psi_2}\cr
    &&Y\cr}\eqno(A.1)$$
commutes, then the diagram between homotopy groups
  $$\matrix{\p{q}(X)&\mapright{\Psi_1}&\p{q}(Z)\cr
    &_{\Phi}\!\searrow\ &\mapdown{\Psi_2}\cr
    &&\p{q}(Y)\cr}\eqno(A.2)$$
will also commute.\footnote{$^5$}{In still other words, we are defining
the homotopy functor from the category of topological spaces to the
category of groups.  See [16].}
(Recall that a diagram is said to commute if, for
any two objects in the diagram and any two maps between the objects,
obtained by composition of maps in the diagram, those two maps
coincide.)  This is straightforward to show, by considering how the
maps between spaces induce maps between their homotopy groups.

Given $\phi:X\rightarrow Y$, we construct $\Phi:\p{q}(X)\rightarrow
\p{q}(Y)$ in the following way.
Fix a map $f_0:S^q\rightarrow X$, which represents
the homotopy class $[f_0]\in\p{q}(X)$.  Then $\phi\circ f_0$ is a map
from $S^q\rightarrow Y$, which represents $[\phi\circ f_0]\in
\p{q}(Y)$. We therefore define $\Phi$ via
$\Phi:[f_0]\mapsto [\phi\circ f_0]$.  We need merely to show
that this definition is independent of our choice
of $f_0$, {\it i.e.}~that it sends two homotopic maps $f_0,\ f_1:
S^q\rightarrow X$ to the same class in $\p{q}(Y)$.  To do this,
consider a homotopy\footnote{$^4$}{Each map must have the same base
point.  That is, $f_0$, $f_1$ and $F$ must all send the north pole
of $S^q$ to the same point in $X$.} from $f_0$ to $f_1$,
given by a map $F:[0,1]\times S^q\rightarrow X$ which satisfies
$F(0)=f_0$, $F(1)=f_1$.  Then the map $\phi\circ F$ serves as a
homotopy from $\phi\circ f_0$ to $\phi\circ f_1$, and hence the map
$\Phi$ is well defined on homotopy classes.

Now consider the composition $\Psi_2\circ\Psi_1$.  $\Psi_1$ takes the
homotopy class of $f:S^q\rightarrow X$ and maps it to the homotopy
class of $\psi_1\circ f:S^q\rightarrow Z$, while $\Psi_2$ takes the
homotopy class of $\tilde f:S^q\rightarrow Z$ to the homotopy class of
$\psi_2\circ \tilde f:S^q\rightarrow Y$.  If we choose $\tilde f=
\psi_1\circ f$, we find that $\Psi_2\circ\Psi_1 :[f]\mapsto
[\psi_2\circ(\psi_1\circ f)]$.  Since composition is associative, we
have shown that $\Psi_2\circ\Psi_1$ is the same map as that induced by
$\psi_2\circ\psi_1$.  Thus, if $\phi=\psi_2\circ\psi_1$, then
$\Phi=\Psi_2\circ\Psi_1$.

\vfill\eject

\centerline{\bf REFERENCES}

\item{1.} N. Turok, {\it Phys. Rev. Lett.} {\bf 63}, 2625 (1989).

\item{2.}
N. Turok and D. Spergel, {\it Phys. Rev. Lett.} {\bf 64}, 2763
(1990);  D. Spergel, N. Turok, W.H. Press, and B.S. Ryden,
{\it Phys. Rev. D} {\bf 43}, 1038 (1991);  R.Y. Cen, J.P.
Ostriker, D.N. Spergel, and N. Turok, {\sl Ap. J} {\bf 393},
42 (1992).

\item{3.} D.P. Bennett and S.H. Rhie, {\it Ap. J. Lett.} {\bf 406},
7 (1993); U. Pen, D.N. Spergel, and N. Turok, {\it Phys. Rev. D},
in press (1993).

\item{4.} I. Chuang, R. Durrer, N. Turok, and B. Yurke,
{\it Science} {\bf 251}, 1336 (1991).

\item{5.} T. Skyrme, {\it Proc. Roy. Soc.} {\bf A262}, 237 (1961).

\item{6.} R. Holman, S.D.H. Hsu, E.W. Kolb, R. Watkins, and
L.M. Widrow, {\it Phys. Rev. Lett} {\bf 69}, 1489 (1992);
M. Kamionkowski and J. March-Russell, {\it Phys. Rev. Lett.}
{\bf 69}, 1485 (1992).

\item{7.} L. Perivolaropoulos, {\it Phys. Rev. D} {\bf 46}, 1858
(1992).

\item{8.} M.J. Greenberg and J.R. Harper {\it Algebraic Topology}
(Benjamin/Cummings Co., Reading, MA, 1981).

\item{9.} G.W. Whitehead, {\it Elements of Homotopy Theory}
(Springer-Verlag, New York, 1978).

\item{10.} S. Coleman, {\it Aspects of Symmetry} (Cambridge Univ.
Press, Cambridge, UK, 1985).

\item{11.} R. Bott, {\it Bull. Soc. Math. France} {\bf 84}, 251
(1956).

\item{12.} L.-F. Li, {\it Phys. Rev D} {\bf 9}, 1723 (1974).

\item{13.} E.J. Weinberg, D. London, and J.L. Rosner, {\it
Nucl. Phys.} {\bf B236}, 90 (1984).

\item{14.} J. Borrill, E.J. Copeland, and A.R. Liddle, {\it Phys.
Lett.} {\bf B258}, 310 (1991); R.A. Leese and T. Prokopec,
{\it Phys. Rev. D} {\bf 44}, 3749 (1991); T. Prokopec, A. Sornborger,
and R.H. Brandenberger, {\it Phys. Rev. D} {\bf 45}, 1971 (1992);
S. Aminneborg, {\it Nucl. Phys.} {\bf B388}, 521 (1992).

\item{15.} S.H. Rhie and D.P. Bennett, preprint UCRL-JC-110560 (1992);
X. Luo, {\it Phys. Lett.} {\bf B287}, 319 (1992).

\item{16.} R. Geroch, {\it Mathematical Physics} (Univ. Chicago Press,
Chicago, 1985).

\vfill\eject

\centerline{\bf Table Captions.}
\vskip .8cm

\noindent Table One.
We list the homotopy groups $\p1$ through $\p3$ for the spheres and
compact Lie groups.  ``Ex. Groups'' refers to the exceptional groups
E$_6$, E$_7$, E$_8$, F$_4$, and G$_2$.

\vskip .6cm

\noindent Table Two.
We list the results of our computation of the homotopy groups of
vacuum manifolds $G/H$ for various choices of $G$ and $H$, as
well as the dimensionality of $G/H$.  The integer $n-l$ is always
taken to be greater than or equal to $l$.

\vfill\eject

\settabs4\columns

\centerline{\bf Table I: Homotopy of Lie Groups and Spheres}

\vskip .5cm
\hrule
\vskip .1cm
\hrule
\vskip .2cm
\+ Space $X$ & $\p1(X)$ & $\p2(X)$ & $\p3(X)$  \cr
\vskip .2cm
\hrule
\vskip .2cm
\+ $S^1$ & \Z & 0 & 0 \cr
\+ $S^2$ & 0 & \Z & \Z \cr
\+ $S^3$ & 0 & 0 & \Z  \cr
\+ $S^{n\geq 4}$ & 0 & 0 & 0 \cr
\+ $\so3$ & \Z$_2$ & 0 & \Z \cr
\+ $\so4$ & \Z$_2$ & 0 & $\Z\oplus\Z$  \cr
\+ $\so{n\geq 5}$ & \Z$_2$ & 0 & \Z\cr
\+ $\su{n\geq 2}$ & 0 & 0 & \Z \cr
\+ ${\rm S}[\u{1}\times\u{1}]$ & $\Z$ & 0 & 0 \cr
\+ ${\rm S}[\u{n}\times\u{1}]$ & $\Z$ & 0 & $\Z$ \cr
\+ ${\rm S}[\u{n}\times\u{m}]$ & $\Z$ & 0 & $\Z\oplus\Z$ \cr
\+ ${\rm Sp}(n\geq 1)$ & 0 & 0 & \Z  \cr
\+ Ex.~Groups & 0 & 0 & \Z \cr

\vskip .2cm
\hrule
\vskip .1cm
\hrule
\vskip .4cm

\vfill\eject

\settabs7\columns

\centerline{\bf Table 2: Homotopy of Vacuum Manifolds}

\vskip .5cm
\hrule
\vskip .1cm
\hrule
\vskip .2cm
\+ $G$ & $H$ && $\p1(G/H)$ & $\p2(G/H)$ & $\p3(G/H)$ & dim($G/H$) \cr
\vskip .2cm
\hrule
\vskip .2cm
\+ $\su{n}$ & 0 && 0 & 0 & \Z & $n^2-1$ \cr
\+ $\su{2}$ & U(1), $\so2$ && 0 & \Z & \Z & $2$ \cr
\+ $\su{n}$ & $\su{n-2}\times\su2$ && 0 & 0 & 0 & $4n-7$ \cr
\+ $\su{n\geq 3}$ & S[$\u{n-l}\times\u{l}$]
     && 0 & \Z & 0 & $2l(n-l)$ \cr
\+ $\su{n\geq3}$ & $\su{n-l}$ && 0 & 0 & 0 & $l(2n-l)$ \cr
\+ $\su{2l[+1]}$ & $\sp{2l}$ && 0 & 0 & 0 & $2l^2-l-1\ [+4l+1]$ \cr
\+ $\su{3}$ & $\so3$ && 0 & \Z$_2$ & \Z$_4$ & $5$ \cr
\+ $\su{n\geq 4}$ & $\so{n}$ && 0 & \Z$_2$ & \Z$_2$
    & ${1\over 2}n(n+1)-1$ \cr
\+ $\so{n\ne 4}$ & 0 && \Z$_2$ & 0 & \Z & ${1\over 2}n(n-1)$ \cr
\+ $\so{4}$ & 0 && \Z$_2$ & 0 & $\Z\oplus \Z$ & $6$ \cr
\+ $\so{n\ne 4}$ & $\so2$ && 0 & \Z & \Z & ${1\over 2}n(n-1)-1$ \cr
\+ $\so{4}$ & $\so2$ && 0 & \Z & $\Z\oplus \Z$ & $5$ \cr
\+ $\so{4}$ & $\so3$ && 0 & 0 & \Z & $3$ \cr
\+ $\so{n\geq 5}$ & $\so3$ && 0 & 0 & \Z$_2$ & ${1\over 2}n(n-1)-3$ \cr
\+ $\so{n\geq 5}$ & $\so{n-l}$ && 0 & 0 & 0 & ${1\over 2}l(2n-l-1)$ \cr

\+ $\so{3}$ & $\so2\times\Z_2$ && $\Z_2$ & \Z & \Z & $2$ \cr
\+ $\so{4}$ & $\so3\times\Z_2$ && $\Z_2$ & 0 & \Z & $3$ \cr
\+ $\so{n\geq 5}$ & $\so{n-1}\times\Z_2$ && $\Z_2$ & 0 & 0 & $n-1$ \cr
\+ $\so{4}$ & $\so2\times\so2\times\Z_2$ && $\Z_2$ & $\Z\oplus\Z$
    & $\Z\oplus\Z$ & 4 \cr
\+ $\so{5}$ & $\so{3}\times\so{2}\times\Z_2$ && $\Z_2$ & $\Z$ & $\Z_2$
    & $6$ \cr
\+ $\so{n\geq 6}$ & $\so{n-2}\times\so{2}\times\Z_2$ && $\Z_2$ & $\Z$ & 0
    & $2(n-2)$ \cr
\+ $\so{6}$ & $\so{3}\times\so{3}\times\Z_2$ && $\Z_2$ & $\Z_2$ & $\Z_2$
    & $9$ \cr
\+ $\so{n\geq 7}$ & $\so{n-l}\times\so{l}\times\Z_2$ && $\Z_2$ & $\Z_2$ & 0
    & $l(n-l)$ \cr
\+ $\so4$ & $\u2$ && $\Z_2$ & 0 & $\Z$ & 2 \cr
\+ $\so{2k[+1]}$ & $\u{k}$ && $\Z_2$ & 0 & 0
    & $k^2-k\ [+2k]$ \cr

\vskip .2cm
\hrule
\vskip .1cm
\hrule
\vskip .4cm

\bye